%% file: paper_apj.tex
\documentclass[iop,twocolumn,numberedappendix]{emulateapj}
\pdfoutput=1

\usepackage{apjfonts}
\usepackage{amsmath,amsfonts,amssymb,mathrsfs,graphicx,color}
\usepackage{tocbibind}
\usepackage[toc,page]{appendix}
\usepackage{ulem}
\usepackage{natbib}

\newcommand{\ie}{{\it i.e.}}

\newcommand{\eg}{{\it e.g.}}

\newcommand{\eq}{Eq.}

\newcommand{\fig}{Fig.}
\newcommand{\Fig}{Fig.}

\newcommand{\Sec}{Section}

\newcommand{\App}{Appendix}

\shorttitle{Cosmogenic Neutrinos Challenge the Cosmic Ray Proton Dip Model}
\shortauthors{Heinze, Boncioli, Bustamante \& Winter}

\begin{document}


\title{Cosmogenic Neutrinos Challenge the Cosmic Ray Proton Dip Model}

\author{Jonas Heinze\altaffilmark{1}, Denise Boncioli\altaffilmark{1}, Mauricio Bustamante\altaffilmark{2,3}, \& Walter Winter\altaffilmark{1}}
\affil{
$^{1}$ Deutsches Elektronen-Synchrotron (DESY), Platanenallee 6, 15738 Zeuthen, Germany\\
$^{2}$ Center for Cosmology and AstroParticle Physics (CCAPP), The Ohio State University, Columbus, OH 43210, USA\\
$^{3}$ Department of Physics, The Ohio State University, Columbus, OH 43210, USA\\
jonas.heinze@desy.de, denise.boncioli@desy.de, bustamanteramirez.1@osu.edu, walter.winter@desy.de
}

\begin{abstract}
The origin and composition of ultra-high-energy cosmic rays (UHECRs) remain a mystery. The proton dip model describes their spectral shape in the energy range above $10^9$~GeV by pair production and photohadronic interactions with the cosmic microwave background. The photohadronic interactions also produce cosmogenic neutrinos peaking around $10^9$ GeV.  We test whether this model is still viable in light of recent UHECR spectrum measurements from the Telescope Array experiment, and upper limits on the cosmogenic neutrino flux from IceCube. While two-parameter fits have been already presented, we perform a full scan of the three main physical model parameters: source redshift evolution, injected proton maximal energy, and spectral index. We find qualitatively different conclusions compared to earlier two-parameter fits in the literature: a mild preference for a maximal energy cutoff at the sources instead of the Greisen--Zatsepin--Kuzmin (GZK) cutoff, hard injection spectra, and strong source evolution. The predicted cosmogenic neutrino flux exceeds the IceCube limit for any parameter combination. As a result, the proton dip model is challenged at more than 95\% C.L. This is strong evidence against this model independent of mass composition measurements.
\end{abstract}

\keywords{Cosmic rays --- Neutrinos --- Astroparticle physics --- Methods: numerical}

\maketitle

\section{Introduction}

Ultra-high-energy cosmic rays (UHECRs) are charged particles of astrophysical origin with energies above $10^9$ GeV, the highest observed. Their sources are unknown, but their energy spectrum has been measured with increasing precision~\citep{Valino:2015,Ivanov:2015}. It exhibits a hardening at about $5 \times 10^9$ GeV -- the ``ankle'' -- and a strong suppression at the topmost energies, around $5 \times 10^{10}$ GeV.

If UHECRs above $10^9$ GeV are mainly protons of extragalactic origin, the spectral features can be attributed to interactions with the cosmic microwave background (CMB) and the infrared/optical photon background (CIB). In this ``proton dip'' model, energy losses due to electron-positron pair production on CMB photons are responsible for the ankle \citep{DeMarco:2003ig,Berezinsky:2005cq,Berezinsky:2002nc,Aloisio:2006wv,Aloisio:2007rc}. The UHECR spectrum is consistent with a power law in energy, with spectral index of $2.4 - 2.8$, where lower values imply a stronger redshift evolution of the number density of UHECR sources \citep{DeMarco:2005ia}. Photopion production on the CMB creates a high-energy cutoff -- the ``GZK cutoff''~\citep{Greisen:1966jv,Zatsepin:1966jv}. This is effectively a cosmic ray horizon: protons detected with energy above the GZK cutoff were necessarily born in the local universe. Photopion interactions also create ``cosmogenic neutrinos'', with $\sim 10^9$ GeV.

An alternative to the proton dip model posits that the transition to a flux dominated by extragalactic cosmic rays occurs at the ankle. Additionally, if UHECRs are a mixture of nuclei, the interpretation of spectral features is more intricate\footnote{For a review on the influence of the extragalactic propagation of UHECRs on their energy spectrum and composition, see~\citet{Allard:2011aa}.}; the flux suppression at the highest energies is due to the photodisintegration of nuclei on the photon backgrounds. Presently, the proton dip, ankle, and mixed composition models all remain ostensibly viable alternatives. We will test whether the former still is.

The largest UHECR observatories -- the Pierre Auger Observatory \citep{ThePierreAuger:2015rma} and the Telescope Array (TA) \citep{AbuZayyad:2012kk} -- aim to settle the issue. They detect UHECR-initiated extensive air showers via surface Cherenkov water tanks or scintillators, fluorescence detectors, or a combination of both techniques. Measurements of the UHECR mass composition, \ie, the relative abundance of lighter versus heavier nuclei, could in principle test the validity of the proton dip model. Composition is determined chiefly by measuring the column depth in the atmosphere at which the particle content of a shower is maximal.

While TA finds consistency with a light primary composition above $10^9$ GeV \citep{Belz:2015}, Auger finds that the mass of the primary reaches a minimum around $10^{9.3}$ GeV before rising with energy \citep{Porcelli:2015}. Measurements of the correlation between the depth of the shower maximum and the number of muons can contribute to the determination of the primary composition. A similar idea is used within the Auger analysis in the energy range of the ankle, finding that the results are in favor of a mixed composition \citep{Yushkov:2015}. Taken at face value, the Auger results could be interpreted as evidence against the proton dip model~\citep{Allard:2005ha,Allard:2008gj,Aloisio:2009sj}. However, the depths of the shower maximum of the two experiments agree within systematic uncertainties \citep{Unger:2015}, and strong conclusions about composition are unattainable due to uncertainties in the interaction of different primaries and shower development \citep{Aab:2014kda,Aab:2014aea}.

A self-consistent interpretation of spectrum and composition results would require propagating a mix of nuclei that interact with the CMB and CIB~\citep{Allard:2008gj,Taylor:2011ta,Fang:2013cba,Aloisio:2013hya,Taylor:2013gga,Taylor:2015rla,Peixoto:2015ava,Globus:2015xga,Unger:2015laa,diMatteo:2015}. We do not attempt to discuss such a mixed-composition interpretation.

In this paper, we instead ask whether the simplest UHECR model -- the proton dip model -- is still viable in light of two recent results: the UHECR spectrum measurements from the TA Collaboration, comprising 7 years of data~\citep{TAcombined,Ivanov:2015}, and the recent upper bound on the flux of cosmogenic neutrinos from the IceCube Collaboration~\citep{Ishihara:2015,icecubeupdate}. We do not use mass composition measurements.

Assuming a population of generic extragalactic proton sources, we scan simultaneously over the three key model parameters: proton spectral injection index, maximal injected energy, and source redshift evolution. While two-parameter fits have been performed before by \citet{DeMarco:2005ia,Ahlers:2010fw,Kido:2015}, this is the first reported three-parameter fit.

We find a compelling conclusion: the high cosmogenic neutrino fluxes implied by TA data challenge the proton dip model at $>95\%$ C.L., for any parameter combination.

This paper is organized as follows. We present our proton injection and propagation model, and the fitting procedure, in \Sec~\ref{sec.propagation}. In \Sec~\ref{sec.fit} we show the results of 2D and 3D scans of the  parameter space. In \Sec~\ref{sec.neu} we calculate the associated cosmogenic neutrinos. We summarize and conclude in \Sec~\ref{sec.conclusions}. \App~\ref{sec.appA} shows the result of the fit performed using the same assumptions related to the sources as TA, while \App~\ref{sec.appB} shows that our conclusions are robust to model variations.

\section{Model and methods}\label{sec.propagation}
\input{propagation.tex}

\section{3D fits of UHECR parameters}\label{sec.fit}
\input{fit.tex}

\section{Cosmogenic neutrinos}\label{sec.neu}
\input{neutrinos.tex}

\section{Summary and conclusions}\label{sec.conclusions}
\input{conclusions.tex} 

\acknowledgments

We thank Dmitri Ivanov and Peter Tinyakov for useful communications; and John Beacom, Shirley Li, Kohta Murase and Alan Watson for valuable discussion. MB was partially supported by NSF Grant PHY-1404311 to JFB. This project has received funding from the European Research Council (ERC) under the European Union's Horizon 2020 research and innovation programme (Grant No.~646623).

\bibliographystyle{apj}

\appendix

\section{Fits using the astrophysical TA assumptions}\label{sec.appA}
\input{appA.tex}

\section{Varying the model and fitting procedure}\label{sec.appB}
\input{appB.tex}

\end{document}

%% file: propagation.tex
 
Our fundamental assumption is that of an extragalactic pure-proton UHECR composition above $10^9$ GeV. The main parameters of our cosmic ray transport model are the spectral injection index $\gamma$, maximal proton energy reached via acceleration at the source $E_{\mathrm{max}}$, and source evolution parameter $m$. We explain them below.

We assume a homogeneous distribution of identical sources with proton injection (in the cosmologically co-moving frame)
\begin{equation}
J^\text{inj}_p(E) \propto \left\{ 
\begin{array}{ll}
H(z) E^{-\gamma} \exp(-E/E_{\mathrm{max}}) & z\ge z_{\mathrm{hom}} \\
0 & z< z_{\mathrm{hom}} \\
\end{array}
\, \right. \, , \label{equ:injflux} 
\end{equation}
{where $E$ is the energy of the injected protons. Here, $z_{\mathrm{hom}}$ is a redshift injection cutoff, below which the (local) universe is inhomogeneous. The dimensionless function $H(z)$ implements the redshift evolution of the number density of sources, normalized by $H(0)=1$.

A frequently used assumption is $H(z) \propto (1+z)^k$, for small $z$. This is appropriate for UHECR propagation dominated by low redshifts. However, higher redshifts $z>1$ significantly contribute to the flux of cosmogenic neutrinos, and the above parameterization could potentially overproduce neutrinos.

We instead parameterize the source evolution relative to the star formation rate (SFR), given by $H_{\mathrm{SFR}}(z)$ in \citet{Hopkins:2006bw}, \ie, $H(z)= (1+z)^m \times H_{\mathrm{SFR}}(z)$, or
\begin{equation}
H(z) = (1+z)^m \times 
\begin{cases} (1+z)^{3.44}, & z\leq 0.97\\ 
10^{1.09} (1+z)^{-0.26}, & 0.97 < z \leq 4.48\\ 
10^{6.66} (1+z)^{-7.8}, & z>4.48  \end{cases} \, .
\label{eq.sourceevo}
\end{equation} 
We simulate injection up to $z = 6$. We allow for positive and negative values of $m$: $m=0$ corresponds to SFR evolution; $m=-3.4$, to no evolution locally; and $m<-3.4$, to negative evolution ~\citep{Taylor:2015rla}. Typically, however, $m>0$ to account for the fact that the luminosity of a source class was higher in the past. Common values are $m \simeq 1.2 \-- 1.4$ for gamma-ray bursts and somewhat larger values (locally) for active galactic nuclei~\citep{Gelmini:2011kg}.

Our parameterization covers, without loss of generality, all possible evolutions for $z \le 1$. This is the region that will dominate our cosmic ray fit above $10^9$ GeV, as $z \sim 1$ roughly corresponds to the maximal proton interaction length in that energy range. However, since a substantial contribution of neutrinos comes from higher redshifts, alternative evolution scenarios for $z > 1$ will affect their flux; see \App~\ref{sec.maxredshift} for a discussion of the most extreme case (no injection for $z > 1$). 

The fits are also affected by the redshift $z_{\mathrm{hom}}$, below which injection is switched off. Since there are no obvious close-by UHECR sources, we know that $z_{\mathrm{hom}}$ must be significantly larger than zero. On the other hand,  $z_{\mathrm{hom}} \lesssim 0.02$, as the universe appears homogeneous at distances beyond the scale of galaxy clusters ($\sim 100$ Mpc). The value of $z_{\mathrm{hom}}$ will affect the cosmic ray propagation above $10^{11} \, \mathrm{GeV}$, where the photohadronic interaction length could become smaller than this scale. A lower value of $E_{\mathrm{max}}$ could be alternatively interpreted as a depletion of local sources, \ie, as a larger value of $z_{\mathrm{hom}}$~\citep{Aloisio:2010wv}. On account of this, and due to limited computation time, we do not consider $z_{\mathrm{hom}}$ as an additional parameter to be fit, but instead fix $z_{\mathrm{hom}} \simeq 0$. 

Additional model parameters could be introduced, such as breaks in the injection spectrum motivated by different escape components from the sources (see, \eg, \Sec~7 in \citet{Baerwald:2013pu}) or alternative shapes of the maximal energy cutoff. We do not consider these, as the discussion would go beyond the scope of this paper.

We compute proton propagation numerically via a transport equation that includes adiabatic, pair production, and photohadronic energy losses; for details, see \App~B in \citet{Baerwald:2014zga}. Photohadronic interactions are computed efficiently, following \citet{Hummer:2010vx}, based on \citet{Mucke:1999yb}. We use the CIB at $z=0$ from \citet{Franceschini:2008tp} and scale it by the SFR at higher redshifts. The effects of magnetic fields are not included, which is a good approximation above $10^9 \, \mathrm{GeV}$~\citep{Aloisio:2006wv}. We adopt a $\Lambda$CDM cosmology with $\Omega_m = 0.27$, $\Omega_\Lambda = 0.73$ and $H_0$ = 70.5 km s$^{-1}$ Mpc$^{-1}$ \citep{Komatsu:2010fb}.

We scan over combinations of $\gamma$, $\log_{10}(E_{\mathrm{max}}/\mathrm{GeV})$, and $m$. For each combination, we fit the computed proton flux at Earth to the latest (7-year) combined -- surface and fluorescence -- energy spectrum measured by the TA Collaboration~\citep{TAcombined,Ivanov:2015}. In \citet{Kido:2015}, the TA Collaboration carried out 2D fits to its surface detector (SD) energy spectrum, which is measured above $10^{9.2}$ GeV. In the present work we fit the TA combined spectrum dominated by SD data, starting from $10^{9.2}$ GeV. This choice reflects that below $10^{9.2}$ GeV different data sets contribute, which means that the spectrum may be affected by several sources of (relative) systematics; a discussion of that is beyond the scope of this paper. Note that  the ankle is fully included in that energy range, and that the transition energy is high enough that magnetic field effects on the cosmic ray propagation can be neglected; see discussion below. We show in \App~\ref{sec.startingene} that the results do not change qualitatively by lowering the starting energy of the fit; however, the fit gets worse due to the data point at $10^{9.15}$ GeV, that is off with respect to the other data points. We have considered the statistical uncertainty of the observed number of events in each energy bin. The systematic uncertainty is dominated by the uncertainty in the energy scale, estimated to be $\sigma_E = 20\%$~\citep{Ivanov:2015}, which we include in the fits as a penalty.

\begin{figure*}[!ht]
\centering
\includegraphics[width=0.8\textwidth]{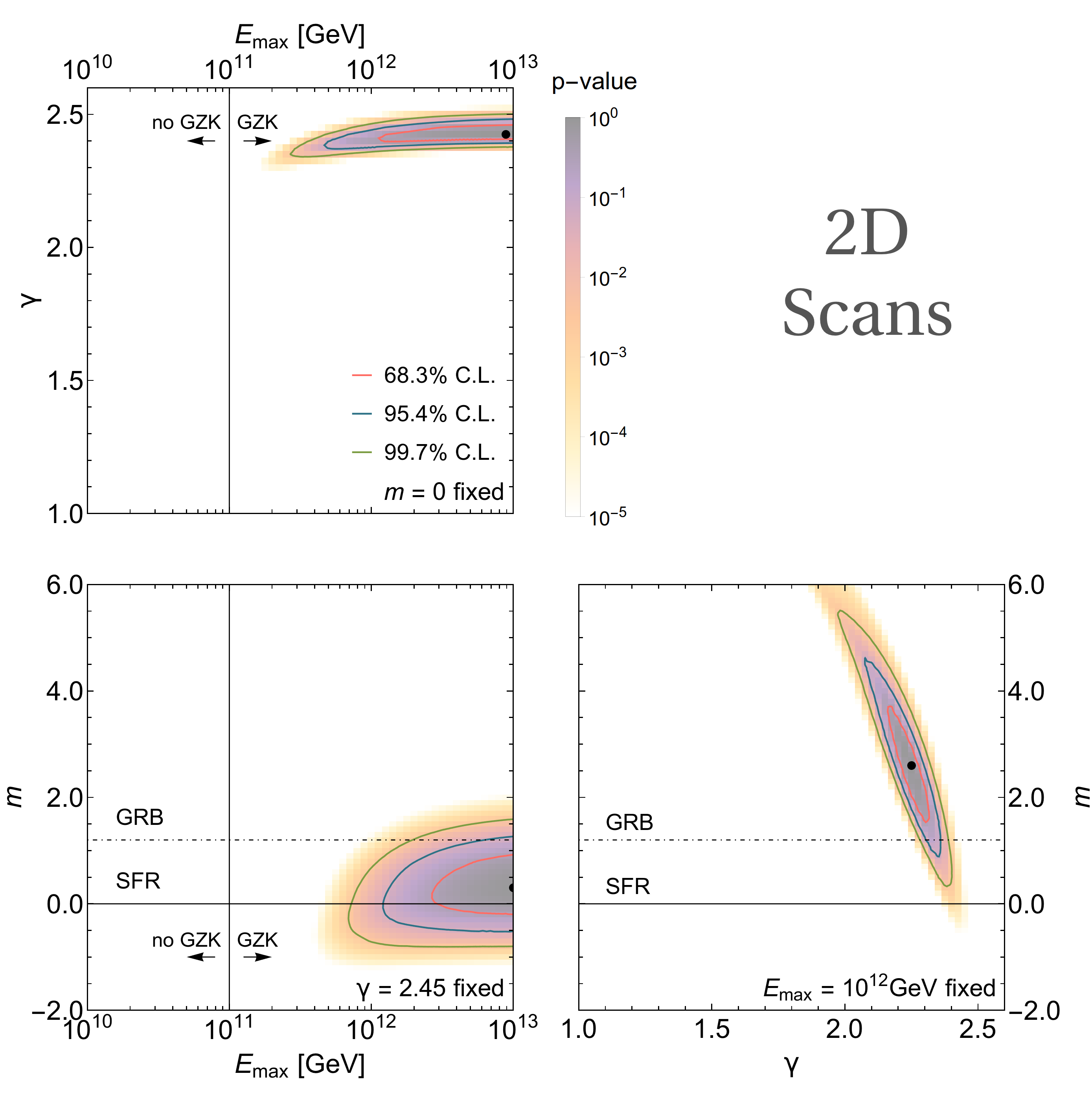}
\caption{Allowed two-parameter regions (2 d.o.f.) from 2D fits to TA spectral data~\citep{TAcombined,Ivanov:2015}. The third parameter is kept fixed to the value shown in each panel. Typical values of $m$ for star formation rate (SFR) evolution and gamma-ray bursts (GRB) evolution are shown for reference.}
\label{fig.2Dscan}
\end{figure*}

For each parameter combination, the free normalization $f$ and the shift in the energy scale $\delta_E$ are found by minimizing the $\chi^2$ estimator
\begin{equation}
\chi^2 = \sum_i \frac{ \left( f J_p^{\mathrm{mod}}(E^{'}_i;\gamma,E_{\mathrm{max}},m)-J_p^{\mathrm{TA}}(E_i) \right)^2} {\sigma^2_i} + \left( \frac{\delta_E}{\sigma_E} \right)^2, 
\label{eq.chi}
\end{equation}
where $J_p^{\mathrm{mod}}(E^{'}_i;\gamma,E_{\mathrm{max}},m)$ is the calculated (model), unnormalized flux at Earth, evaluated at the shifted energy $E^{'}_i \equiv (1+\delta_E)E_i$, and $J_p^{\mathrm{TA}}(E_i)$ is the measured flux at energy $E_i$. At $E_i$, the statistical error is $\sigma_i$. The sum is performed over the 21 TA data points above $10^{9.2}$ GeV. The last term is a penalty from systematics; see \App~\ref{sec.syserror} for further discussion of systematics.
The best fit is obtained at minimum $\chi^2$, \ie,
\begin{equation}
\chi^2_{\mathrm{min}} = \underset{f,\delta_E;\gamma,E_{\mathrm{max}},m}{\text{Min}} \;  \chi^2(f,\delta_E;\gamma,E_{\mathrm{max}},m) \, ,
\end{equation}
which indicates the goodness of fit, whereas $\Delta \chi^2 = \chi^2 - \chi^2_{\mathrm{min}}$ determines the allowed regions. Both $f$ and $\delta_E$ are considered nuisance parameters, while the physical parameters are $\gamma$, $E_\text{max}$, and $m$.

Our main result is a 3D scan over $\gamma$, $E_\text{max}$, and $m$. We will show it as 2D projections via
\begin{equation}
 \Delta \chi^2(a,b) = \underset{f,\delta_E;c}{\text{Min}} \; \Delta \chi^2(f,\delta_E;\gamma,E_{\mathrm{max}},m) \, , \label{equ:3d}
\end{equation}
for 2 d.o.f. Here, $a$, $b$, and $c$ are three different parameters from the set $\{\gamma,E_{\mathrm{max}},m \}$ and $\Delta \chi^2 (a,b)$ is minimized over $c$. In some cases, we will show 2D scans to compare to the existing literature; in these, one of the parameters will be fixed for the whole procedure, including the computation of $\chi^2_{\mathrm{min}}$.

%% file: fit.tex
\begin{figure*}[!ht]
\centering
\includegraphics[width=0.8\textwidth]{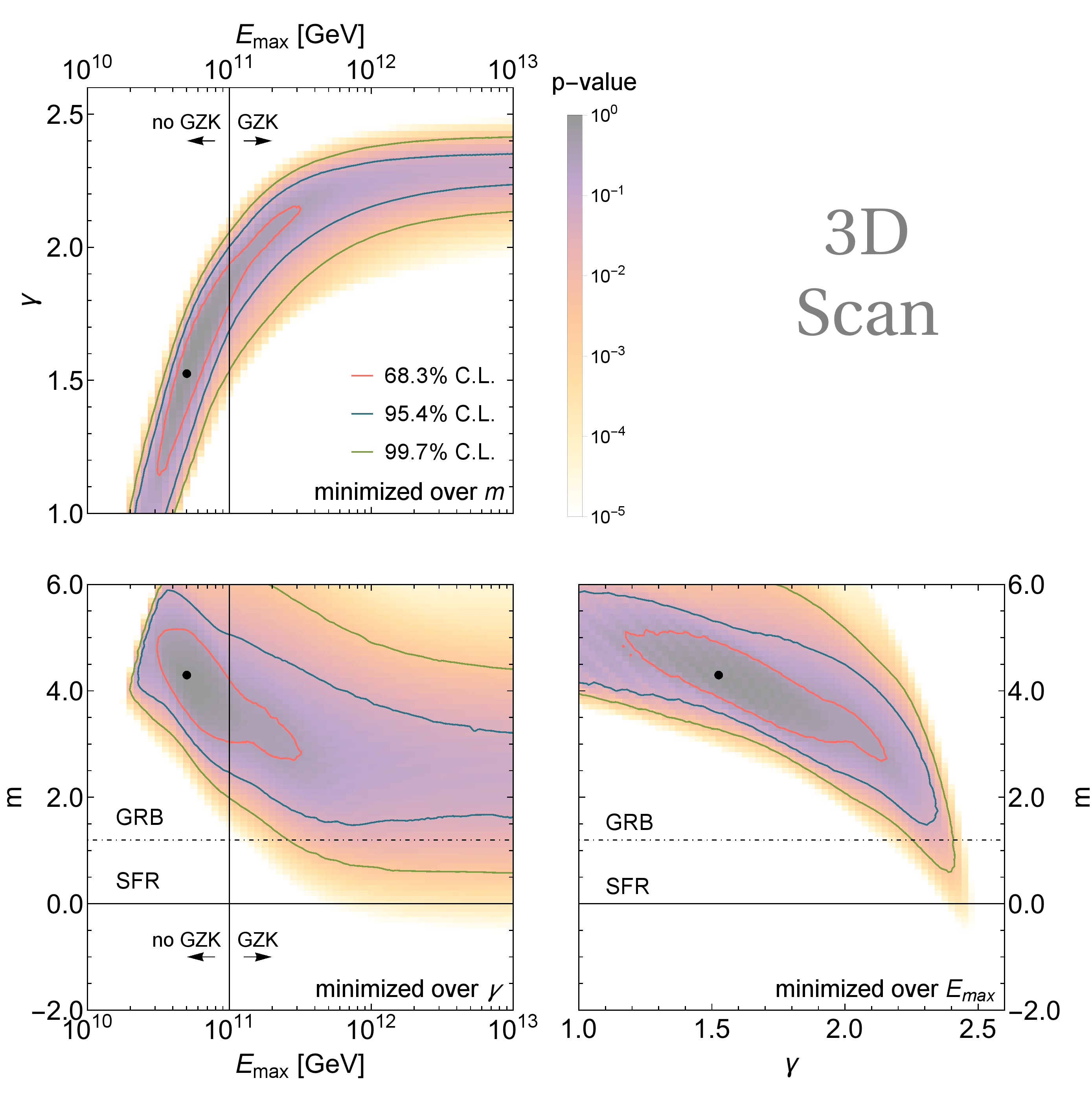}
\caption{Allowed two-parameter regions (2 d.o.f.) from a 3D fit to TA spectral data~\citep{TAcombined,Ivanov:2015}. In each panel, the third parameter has been minimized over.}
\label{fig.3Dscan}
\end{figure*}

\begin{table}[t!]
 \begin{center}
  \caption{Best-fit parameters, 1$\sigma$ uncertainties (for the 3D scan only), and $\chi^2_{\mathrm{min}}/{\mathrm{d.o.f.}}$ for the 2D scans and for the 3D scan. The cases marked with ${}^*$ refer to the fixed parameter in the 2D scans. \label{tab.bestpar}}
  \begin{tabular}{ccccc}
   \tableline
   \tableline
   \multicolumn{1}{c}{\textbf{}}&
   \multicolumn{3}{c}{\textbf{2D scans}}&
   \multicolumn{1}{c}{\textbf{3D scan}}\\
   \tableline
   \vspace{-0.2cm}\\
   $\gamma$ & 2.25 & ${}^*2.45$  & 2.42 & $1.52^{+0.35}_{-0.20}$\\[5pt]
   $\log_{10}(E_{\mathrm{max}}/\mathrm{GeV})$ & ${}^*12.0$  & $13.0$ & 12.9 & $10.7^{+0.3}_{-0.1}$\\[5pt]
   $m$ & $2.6$ & 0.3 & ${}^*0.0$ &$ 4.3^{+0.4}_{-0.8}$\\[5pt]
   $\delta_E$ & $-0.11$& $-0.06$ & $-0.16$ & $-0.35$\\[5pt] 
   $\chi^2_{\mathrm{min}}/{\mathrm{d.o.f.}}$ &$34.7/17$ & $47.8/17$ &$47.8/17$&  $30.8/16$\\[3pt]
   \tableline
  \end{tabular}
 \end{center}
\end{table}

We first perform 2D scans, fixing one of the physical parameters, to compare our results with the existing literature. We show in \App~\ref{sec.appA} that we can reproduce the fits performed by the TA Collaboration \citep{Kido:2015}.

Figure \ref{fig.2Dscan}, lower right panel, shows the result of a fit with fixed $E_{\mathrm{max}} = 10^{12}$ GeV, as in \citet{Kido:2015}, with which it agrees. In the lower left panel, we instead fix $\gamma=2.45$. In the upper panel, we fix $m=0$, corresponding to SFR evolution. The corresponding best-fit parameters are reported in Table \ref{tab.bestpar}.

The regions are relatively small and compatible with similar results in the literature~\citep{DeMarco:2005ia,Ahlers:2010fw,Kido:2015}. Large $E_{\mathrm{max}}$ and soft injection spectra are preferred. This is by construction, since we chose $E_{\mathrm{max}}$ large in the lower right panel, and the other fixed values were picked from it. We will see this change in the 3D scan.

Figure \ref{fig.3Dscan} shows our novel results for the 3D scan in $(\gamma,E_{\mathrm{max}},m)$, as projections onto three different planes using Eq.~(\ref{equ:3d}). Compared to \fig~\ref{fig.2Dscan}, the regions are larger due to multi-parameter correlations. The result is qualitatively different from the 2D scans and from previous literature. The best-fit values of the parameters and their uncertainties are reported in Table \ref{tab.bestpar}. The value of  $\chi^2_{\mathrm{min}}/{\mathrm{d.o.f.}}=30.8/16$ is an improvement over our 2D scans.

The 3D fit slightly prefers lower maximal energies, harder spectra, and stronger source evolution.  The high value of $m$ implies that the contribution of distant sources is enhanced with respect to SFR evolution. The interpretation of the flux suppression at the highest energies as due to the maximal injected energy at the sources (or a $z_{\mathrm{hom}}$ cutoff, or a combination of them) is slightly favored over the GZK cutoff interpretation. However, the $\chi^2$ function is relatively flat and extended in parameter space, so the preference is mild.

The energy scale shift is comparatively large at the best fit, which is in perfect consistency with the chosen statistical procedure.  We discuss in \App~\ref{sec.enescale} the result assuming a fixed energy scale ($\delta_E = 0$), which may be indicative for results with improved energy resolution.

In the 2D scans of \fig~\ref{fig.2Dscan}, one parameter was fixed in each panel. By including it in the fit, the allowed regions are extended. They maintain their flat behavior with $E_\text{max}$ above $\sim 3\times 10^{11}$ GeV. The comparison between the 3D scan minimized over $E_{\mathrm{max}}$ and the 2D scan with fixed $E_{\mathrm{max}} = 10^{12}$ GeV shows that this choice for the maximal energy restricts the allowed parameter space $(\gamma,m)$ to a small region where the fit can only select a combination of high values for $\gamma$ and intermediate-high values for $m$ (lower right panels). By letting $E_\text{max}$ float, many more combinations of $\gamma$ and $m$ become accessible. The spectrum hardens, while the source evolution increases in order to enhance the contribution of distant sources. This happens at the expense of the maximal energy, whose allowed region now extends below the GZK cutoff.

\begin{figure}
\centering
\includegraphics[width=.47\textwidth]{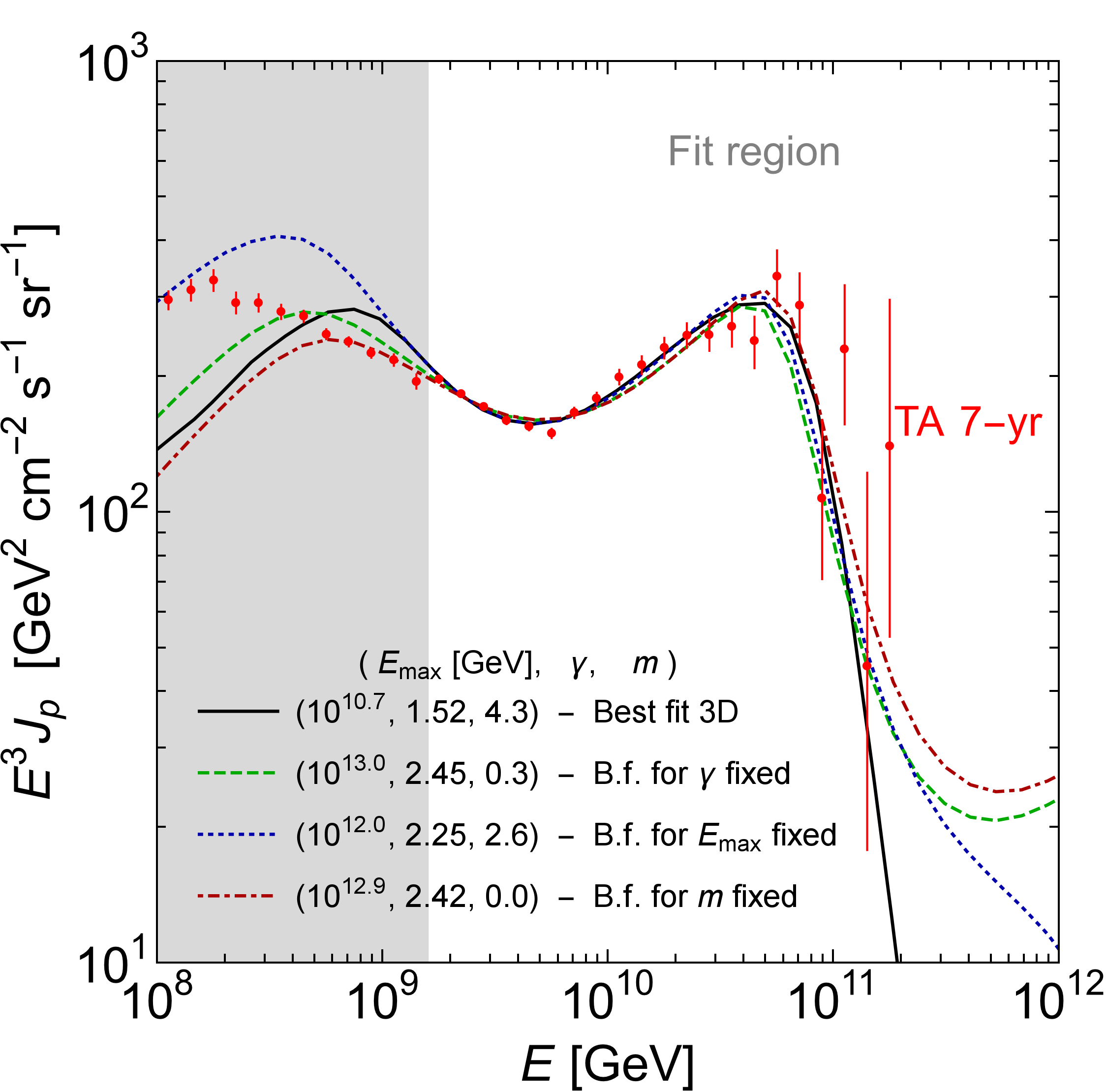}
\caption{Best-fit UHECR spectra for 3D scan (solid curve) and 2D scans (dashed/dotted curves), superimposed on the TA 7-year data ~\citep{TAcombined,Ivanov:2015}. The energy scale of the data points is fixed, while that of the models is for each one shifted by the best-fit value of $\delta_E$.}
\label{fig.protonspectra}
\end{figure}

Figure \ref{fig.protonspectra} shows the best-fit proton spectra at Earth, for the 3D scan (solid) and the 2D scans (dashed/dotted). The energy scale of each spectrum has been shifted by $\delta_E$ resulting from its own fit. At the highest energies, the fluxes for the 2D scans look rather similar because the maximal energy in these cases is larger than the GZK threshold. In contrast, the maximal energy found in the 3D scan is at the level of the GZK threshold, which creates the sharper cutoff at the highest energies, by demanding a large shift in the energy scale.

The effect of source evolution can be clearly seen in the overshooting of the data below $10^{9.2}$ GeV. It is largest for the 2D scan with fixed $E_\text{max}$ because of the combination of soft spectral index and non-zero evolution. In the 3D scan, the evolution is stronger, but the harder spectral index allows the low-energy flux to be smaller than in the 2D scan. 

The low-energy overshooting cannot be compensated and potentially leads to problems elsewhere (\eg, injects too much energy into electromagnetic cascades; see below). However, it can be avoided by different physical mechanisms, depending on the source and propagation models. The propagation of protons in a homogeneous, turbulent extragalactic magnetic field affects their observed spectrum below $10^{9}$ GeV; see, \eg, \citet{Globus:2007bi}. A dependence of the pure proton spectrum below $10^{8}$ GeV on the spatial distribution of the field is found by using more realistic magnetic field configurations (\eg, \citet{Kotera:2007ca}). Several mechanisms have been considered to reduce the flux at low energies. \citet{Berezinsky:2002nc} proposed a change in $\gamma$ below $10^9$ GeV. \citet{Kachelriess:2005xh} proposed a distribution of $E_\text{max}$ to reconcile the measured spectrum with Fermi shock acceleration, with the added bonus of reducing the low-energy flux. \citet{Aloisio:2004fz} discussed an anti-GZK effect related, in the case of diffusive propagation, to an increase of the maximum distance from which UHE protons can arrive. \citet{Lemoine:2004uw} found a low-energy steepening of the spectrum to be a signature of extragalactic magnetic fields; it is used to cut off the extragalactic contribution at the lowest energies. To avoid the overshooting, magnetic field effects on cosmic ray propagation and diffusion \citep{Aloisio:2006wv} can be included, which reduce the low-energy contribution, or a minimal energy cutoff at the sources can be considered -- which we illustrate in \App~\ref{sec.mininj}. In \App~\ref{sec.oversh}, we also show a fit where the overshooting is penalized (assuming that there is no such physical suppression effect), and we find that the resulting astrophysical interpretation is similar to that of our reference case. Since the interpretation of the overshooting effect is model-dependent (which is the beyond the scope of this work), we do not include it in our baseline case.

\begin{figure}
\centering
\includegraphics[width=.47\textwidth]{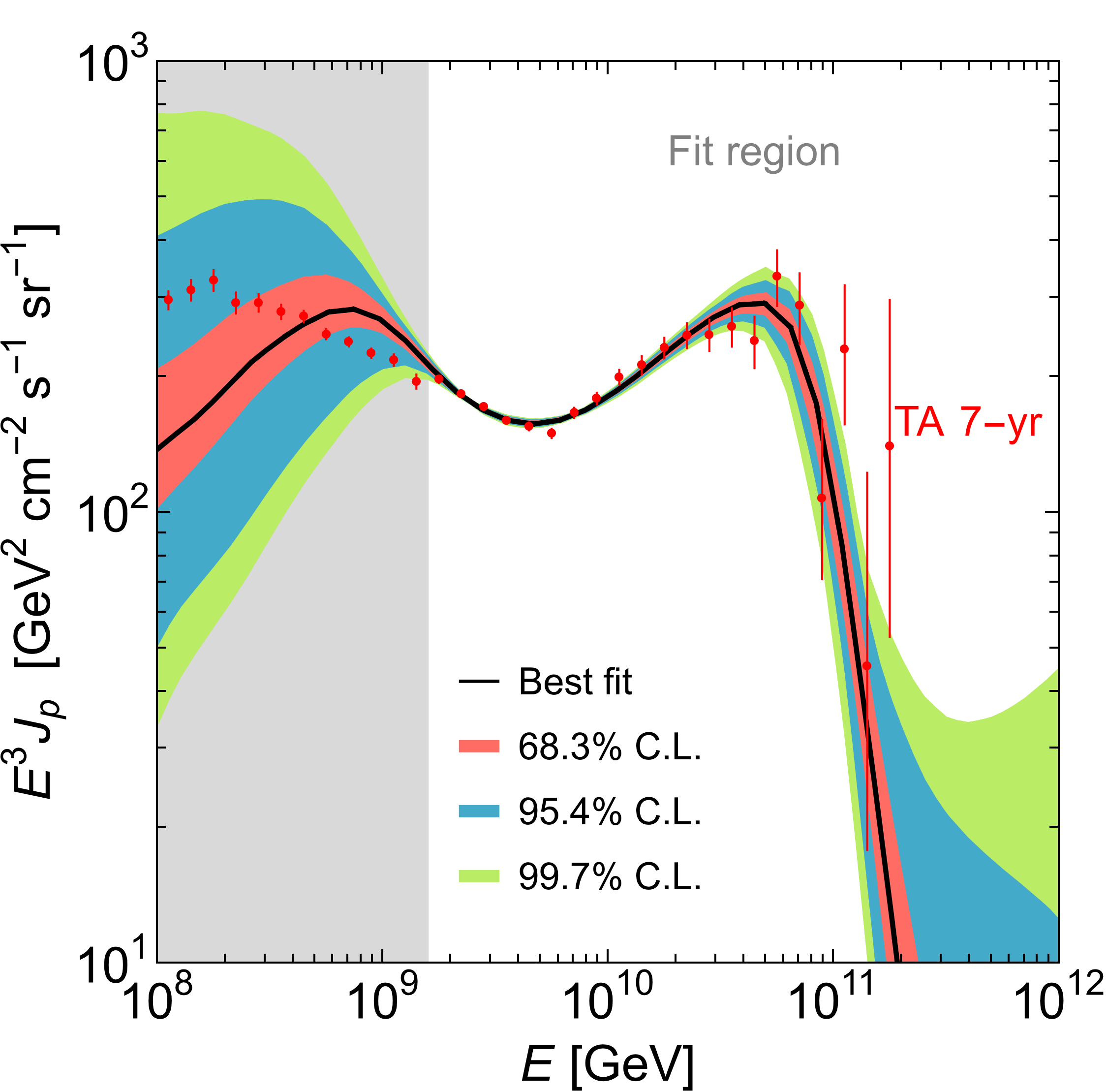}
\caption{Best-fit UHECR spectra for 3D scan (black curve), superimposed on the TA 7-year data ~\citep{TAcombined,Ivanov:2015}, together with the bands determined by the minimal and maximal proton spectra corresponding to each confidence level (at 1 d.o.f.) for each energy.}
\label{fig.protonspectra_range}
\end{figure}

Figure \ref{fig.protonspectra_range} shows the best-fit proton spectrum at Earth and its confidence intervals (at 1 d.o.f.) for the 3D scan. The minimal and maximal proton spectra at each energy are set by the lower and upper edges of the shaded regions at each energy. The largest variations can be seen below the fitting region. The maximal proton flux in this region has to be ascribed to the highest allowed values for $m$ within the 99.7\% C.L., that are connected with the lowest allowed values for $E_\text{max}$, responsible for the minimal proton flux at the highest energies. The maximal proton flux at the highest energies instead corresponds to a larger $E_\text{max}$, which is related to a smaller $m$ that is responsible for the minimal proton flux at the lowest energies.  

For comparison, we performed a fit to the Auger spectrum reported in \citet{Schulz:2013}, assuming a pure proton injection and ignoring the Auger results on composition. A maximal energy cutoff is again preferred over the GZK cutoff. The shift of the energy scale is found to be in the same direction as in the TA fit, though with a smaller value. However, the quality of the fit is worse compared with the TA case and, in fact, unreasonably large values of source evolution are preferred.

%% file: neutrinos.tex
Cosmogenic neutrinos are produced by the decay of pions, muons, kaons, and neutrons produced in photohadronic interactions during cosmic-ray propagation. Unlike protons or nuclei, neutrinos created at high redshifts reach Earth, since they rarely interact and only undergo adiabatic energy losses and flavor mixing. Their flux depends on the composition, production, and propagation of UHECRs~\citep{Hill:1983xs,Engel:2001hd,Kalashev:2002kx,Semikoz:2003wv,Allard:2006mv,Takami:2007pp,Ahlers:2009rf,Ahlers:2010fw,Berezinsky:2010xa,Ahlers:2011jj,Decerprit:2011qe,Gelmini:2011kg,Murase:2012df,Yoshida:2012gf,Roulet:2012rv,Stanev:2014asa,Aloisio:2015ega}. It is strongly affected by source evolution, especially at high redshifts~\citep{Gelmini:2011kg,Aloisio:2015ega}.

If UHECRs are dominated by protons, cosmogenic neutrinos will reach EeV energies; if UHECRs are dominated by nuclei, photomeson production will be less efficient~\citep{Murase:2010gj,Ahlers:2012rz}. A comparison between the expected and measured neutrino fluxes has, in principle, the power to distinguish between the two possibilities.

Ultra-high-energy astrophysical neutrinos, above the PeV scale, are searched for in dedicated experiments \citep{Aslanides:1999vq,Kestel:2004ep,Miocinovic:2005jh,Aggouras:2004mh,Arnold:2004xq,Silvestri:2005qe,Allison:2011wk,Barwick:2014pca} and in experiments designed to detect UHECRs~\citep{Rubtsov:2013,Aab:2015kma,Bleve:2015,Adams:2013nha}. The IceCube Collaboration has detected neutrinos up to a few PeV~\citep{Aartsen:2013bka,Aartsen:2013jdh,Aartsen:2013eka,Aartsen:2014gkd,Aartsen:2015rwa,Aartsen:2015ita}, likely produced on-site at unidentified astrophysical sources. Cosmogenic neutrinos remain undetected, but IceCube searches have recently reached the lowest flux sensitivities~\citep{Ishihara:2015,icecubeupdate}.

\begin{figure}
\centering
\includegraphics[width=.48\textwidth]{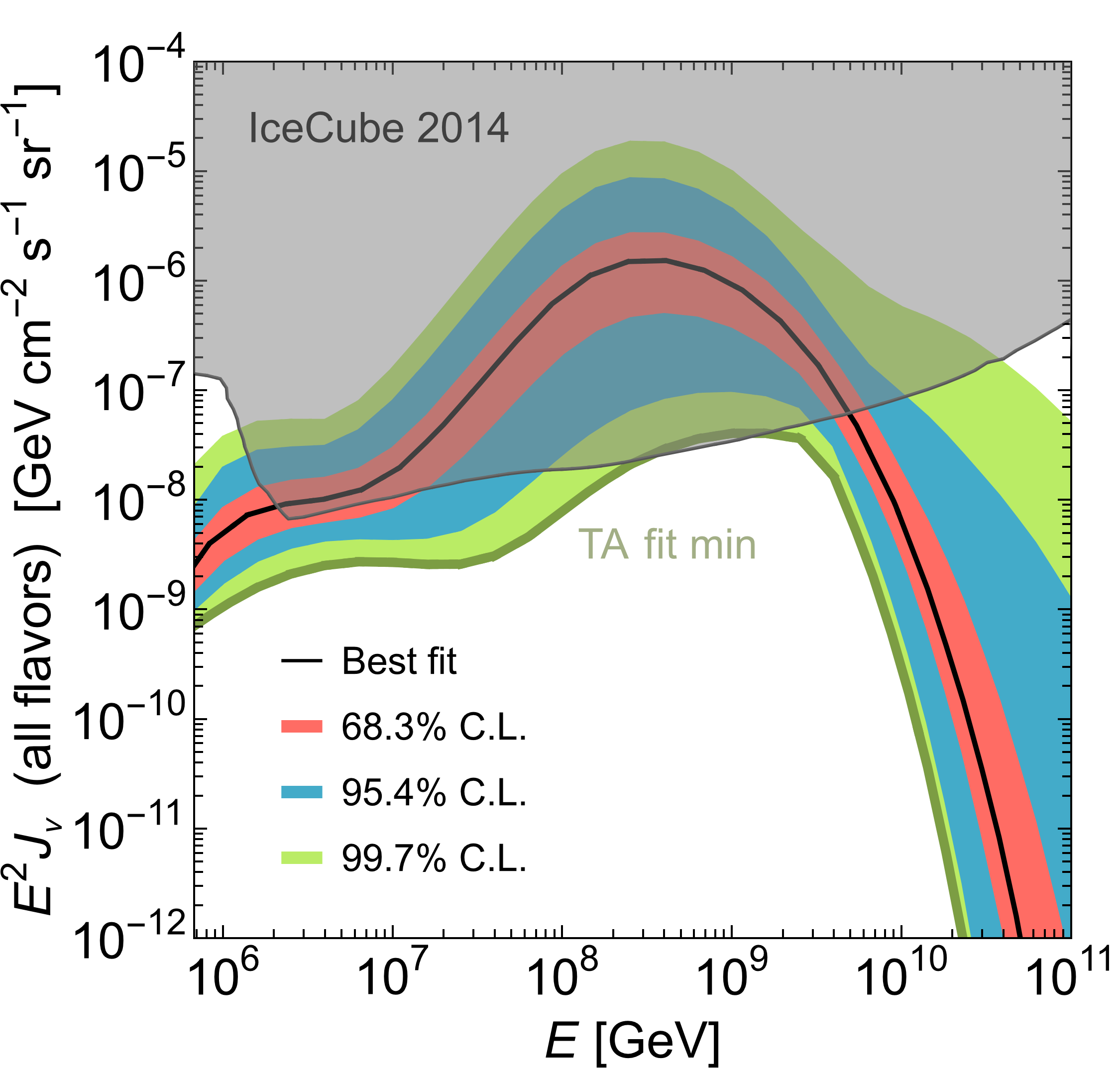}
\caption{All-flavor flux of cosmogenic neutrinos predicted by the 3D fit to the TA 7-year UHECR spectrum reported in \Sec~\ref{sec.fit}. The IceCube upper limit is from \citet{icecubeupdate}.}
\label{fig.neulimit}
\end{figure}

Figure~\ref{fig.neulimit} shows the flux of cosmogenic neutrinos associated to the best-fit proton spectrum obtained in the 3D scan discussed in \Sec~\ref{sec.fit}. The shaded bands around it denote confidence intervals (at 1 d.o.f.) corresponding to the cosmic ray fit. Their edges are obtained by finding, at each energy, the minimal and maximal allowed neutrino fluxes within the chosen confidence level, \ie, they are piece-wise dominated by different neutrino spectra.

The bottom curve -- marked ``TA fit min'' -- is the envelope of all possible neutrino fluxes allowed by the cosmic ray fit. It is in tension with the TA UHECR data at 99.7\% C.L. Below $\sim 4 \times 10^9 \, \mathrm{GeV}$, it corresponds to a larger $E_{\mathrm{max}}$ and a smaller $m$; above, to a smaller $E_{\mathrm{max}}$ and a larger $m$. The proton dip model is disfavored because this minimal envelope exceeds the IceCube upper limit \citep{icecubeupdate}.

The number of expected neutrino events for a given flux $J_\nu(E)$ is given by 
\begin{equation}
 N_\nu = \Delta \Omega \cdot t_{\mathrm{exp}} \int dE \: J_{\nu}(E) \cdot A_{\mathrm{eff}}(E) \; ,
 \label{eq.nnumb}
\end{equation}
where $A_{\mathrm{eff}}(E)$ is the (declination-averaged) effective area for the chosen neutrino flavor(s) including threshold and Earth matter effects, $t_{\mathrm{exp}}$ is the time of exposure, and $\Delta \Omega$ is the solid angle coverage used for the analysis. If no significant flux is observed, the number $N_\nu$ can be interpreted in terms of the confidence level of a limit~\citep{Feldman:1997qc}. Conversely, one often shows that result in terms of a differential upper limit $C/(\Delta \Omega \, E \, t_{\mathrm{exp}} \, A_{\mathrm{eff}}(E))$, where $C$ is a normalization constant depending on confidence level and definition. We estimate the number of expected neutrino events in IceCube from the differential upper limit  given in \citet{icecubeupdate}, where we obtained the normalization constant $C \approx 2.5$ by cross checking the benchmark models \citep{Yoshida:1993pt,Kotera:2010yn,Ahlers:2012rz} shown in \citet{icecubeupdate} with the corresponding event rates.

Table~\ref{tab.neuevents} shows the event rates calculated using the fluxes resulting from our scans. The number of expected events associated to the best-fit solution is more than 20 times what has been shown to be the highest expectation for IceCube \citep{Ahlers:2012rz}, \ie, it can clearly be ruled out. Most importantly, ``TA fit min'' yields 4.9 events. Since only one event was observed, this flux can be excluded at $95\%$ C.L.~\citep{Feldman:1997qc}. Given that this is an unphysical, envelope solution, the actual number of expected events from it is even somewhat higher.

Alternative sets of assumptions are studied in \App~\ref{sec.appB}. They give qualitatively similar results; some of them lead to even stronger conclusions. The only exception, from the point of view of astrophysical assumptions, is if the injection is cut off for $z \gtrsim 1$ (see \App~\ref{sec.maxredshift}), a situation which cannot be identified with UHECRs, since they are almost insensitive to large redshifts. In that case, the cosmogenic neutrino flux would be significantly reduced, and a factor of five larger statistics would be needed to reach the same conclusions as above. However, this extreme scenario might be unrealistic: switching off UHECR injection at $z \gtrsim 1$ seems to contradict the fact that the star formation activity is highest there, and might have consequences elsewhere. We also discussed in \App~\ref{sec.syserror} the results after considering an uncorrelated bin-to-bin systematic error, showing that this treatment of the systematics could reduce the minimal neutrino flux.

\begin{table}[t!]
 \begin{center}
  \caption{Expected number of cosmogenic neutrino events after 6 years in IceCube, corresponding to the 7-year UHECR TA best-fit, and to the minimal fluxes within the 68.3\%, 95.4\%, 99.7$\%$ C.L.\label{tab.neuevents}}
  \begin{tabular}{ll}
   \tableline
   \tableline
   \multicolumn{1}{c}{\textbf{}}&
   \multicolumn{1}{c}{\textbf{$\nu$ events}}\\
   \tableline
   Best fit &  180.6\\
   68.3$\%$ C.L. min flux & 62.7\\
   95.4$\%$ C.L. min flux & 12.4\\
   99.7$\%$ C.L. min flux, {\bf TA fit min}& 4.9\\
   \tableline
  \end{tabular}
 \end{center}
\end{table}

Gamma rays from the decay of $\pi^0$ produced in photohadronic interactions provide a different handle on the cosmic ray injection. They are reprocessed in electromagnetic cascades and, if over-produced, may  overshoot recent Fermi-LAT bounds on the extragalactic gamma-ray background \citep{Ackermann:2014usa}; see \citet{Ahlers:2010fw,Ahlers:2011jj,Berezinsky:2010xa} for discussion.
Since the production of  $\pi^0$ and $\pi^\pm$ are closely correlated, so are the injections into electromagnetic cascades and cosmogenic neutrinos. Therefore, the diffuse gamma-ray data can help constrain the maximal allowed neutrino flux and the corresponding parameter space. Parameter sets leading to low neutrino flux are typically not affected by the Fermi bound. While we do not take the gamma-ray constraint into account explicitly in this work, we note that, since we are interested in the minimal allowed neutrino flux, we do not expect our conclusions to be significantly affected if this additional constraint were imposed. However, the fit regions in \fig~\ref{fig.3Dscan} may be significantly reduced where large neutrino fluxes are produced.

%% file: conclusions.tex
The features of the UHECR energy spectrum are known to high precision, but their origin remains a mystery. The unprecedented sensitivity to the predicted flux of cosmogenic neutrinos can be used as a tool to solve the mystery.

In this work, we have tested the cosmic ray proton dip model, in which UHECRs above $10^9$ GeV are mainly protons of extragalactic origin. We have used the UHECR spectrum recently reported by the Telescope Array using 7 years of data~\citep{TAcombined,Ivanov:2015} and the recent upper limit on cosmogenic neutrinos reported by IceCube~\citep{Ishihara:2015,icecubeupdate}.

We have performed a 3D parameter space scan in terms of spectral injection index, maximal proton energy, and source redshift evolution. The fit to TA data has qualitatively different features compared to 2D scans previously performed in the literature, due to multi-parameter correlations. An interpretation of the data in terms of hard spectra, strong source evolution, and low maximal proton energy is slightly favored over the conventional GZK cutoff scenario -- at the expense of a large systematic shift of the energy scale. 

We have also computed the associated cosmogenic neutrino fluxes in the 3D scan. We have identified the minimal allowed neutrino flux (``TA fit min''), corresponding to the 99.7\% C.L. region allowed by the fit to cosmic ray data. It is in tension with the IceCube upper limit at more than 95\% C.L.

As a result, the conventional proton dip model is challenged for any possible parameter combination of the 3D scan. Our result is a test of the proton dip model completely independent from composition data.

We have also shown the robustness of our results with different sets of assumptions. While some caveats can come from the treatment of the systematics in the fit procedure, one related to the astrophysics of sources is an injection cutoff at $z \gtrsim 1$, which leads to lower neutrino fluxes but hardly affects UHECRs. The corresponding minimal neutrino flux would require about five times more statistics for detection, which should be within reach of the volume upgrade IceCube-Gen2 \citep{Aartsen:2014njl}. Detection of this flux -- though challenging -- in combination with verification of proton composition at the highest energies would be a unique test of cosmic-ray injection beyond the local universe. 

Our result implies that the dip in the cosmic ray spectrum cannot come from pair production in a pure proton model. An obvious interpretation is that the composition of cosmic rays is heavier than protons at the highest energies, which the Auger composition measurements indicate \citep{Porcelli:2015}. Alternatively, the transition to a flux dominated by extragalactic cosmic rays could occur at the ankle energy or higher, while the highest energies are still proton-dominated. We have tested that it is not possible to exclude that ``ankle model'' by simply repeating our analysis  restricting it to the highest-energy data points: in that case, the source evolution -- that is mainly responsible for the excess in the neutrino flux -- is not well-constrained, since at the highest energies only the local universe is responsible for the cosmic ray flux.

%% file: appA.tex
\begin{table*}[b!]
 \begin{center}
  \caption{Best-fit parameters, $\chi^2_{\mathrm{min}}/{\mathrm{d.o.f.}}$, number of expected neutrino events in IceCube after 6 years for the TA best-fit, and the minimal fluxes within the 68.3, 95.4, 99.7$\%$ C.L fit regions of the cosmic ray data. The different rows correspond to different analysis assumptions. The cases marked with ${}^*$ refer to the fixed parameter; the case marked with ${}^{**}$ refers to the dedicated scan assuming the TA assumption for the source evolution.\label{tab.alt}}
  \begin{tabular}{lrrrrr|rrrr}
   \multicolumn{1}{}{\textbf{}} &
   \multicolumn{5}{r|}{\textbf{Best-fit parameters}} &
   \multicolumn{4}{r}{\textbf{Expected $\nu$ events}} \\
   \tableline
   \multicolumn{1}{l}{\textbf{Analysis}} &
   \multicolumn{1}{r}{\textbf{$\gamma$}} &
   \multicolumn{1}{r}{\tiny{\textbf{$\log_{10}(E_{\mathrm{max}}/\mathrm{GeV})$}}} &
   \multicolumn{1}{r}{\textbf{$m$}} &
   \multicolumn{1}{r}{\textbf{$\delta_E$}} &
   \multicolumn{1}{r|}{\textbf{$\chi_{\mathrm{min}}^2/{\mathrm{d.o.f.}}$}} &
   \multicolumn{1}{r}{\textbf{Best-fit}} & 
   \multicolumn{1}{r}{\textbf{68.3\% C.L}} & 
   \multicolumn{1}{r}{\textbf{95.4\% C.L.}} & 
   \multicolumn{1}{r}{\textbf{99.7\% C.L.}} \\ 
   \tableline
   Standard (3D) &1.52 &10.7&4.3& $-0.35$ &30.8/16&180.6 & 62.7 & 12.4 & {\bf 4.9} \\
   2D fit with TA assumptions & 2.22 &${}^*$12.0&${}^{**}$6.5& $-0.10$ &34.2/17&27.8 & 12.9 & 6.9 & {\bf 4.4} \\
   3D fit with fixed energy scale & 2.30 &13.0&2.9& ${}^{*}0.00$ & 39.4/17&40.5 & 17.8 & 7.8 & {\bf 4.4} \\
   3D fit with 3\% syst. added in quadrature & 1.75 &10.8&3.8& $-0.31$ & 22.4/16&94.0 & 30.0 & 5.0 & {\bf 1.5} \\
   3D fit with low-energy penalty & 1.00 &10.3& 4.1 & $-0.51$ &37.8/16&145.1 & 124.4 & 74.4 & {\bf 12.1} \\
   3D fit starting from $10^{9}$ GeV & 1.05 &10.4&4.3& $-0.49$& 40.2/18 & 173.4 & 111.0 & 58.8 & {\bf 2.5} \\
   3D fit with $z=1$ cutoff & 1.90 &11.0& 3.4& $-0.29$&32.3/16&2.5 & 1.9 & 1.3 & {\bf 0.9} \\
   \tableline
  \end{tabular}
 \end{center}
\end{table*}

\citet{Kido:2015} performed a fit to the UHECR energy spectrum measured by the surface detector (SD) array of the Telescope Array after five years of data-taking, assuming a pure proton composition. They used a source evolution of the form $(1+z)^m$ and injected protons up to $z=2$, with fixed $E_{\mathrm{max}} = 10^{12}$ GeV. Their scan of the two-dimensional space $(\gamma,m)$, assuming a uniform distribution of sources, found a best fit at $(\gamma,m)= (2.21^{+0.10}_{-0.15}, 6.7^{+1.7}_{-1.4})$, for $\delta_E=-0.03$, with $\chi_{\text{min}}^2/{\mathrm{d.o.f.}}=12.4/17$.

We report here our results adopting the same choices for the source evolution function and fixed maximal energy. The best-fit parameters of this dedicated two-dimensional scan are repoted in Table \ref{tab.alt} (second row).

\begin{figure*}[t!]
\begin{tabular}{l l}
\includegraphics[width=0.45\textwidth]{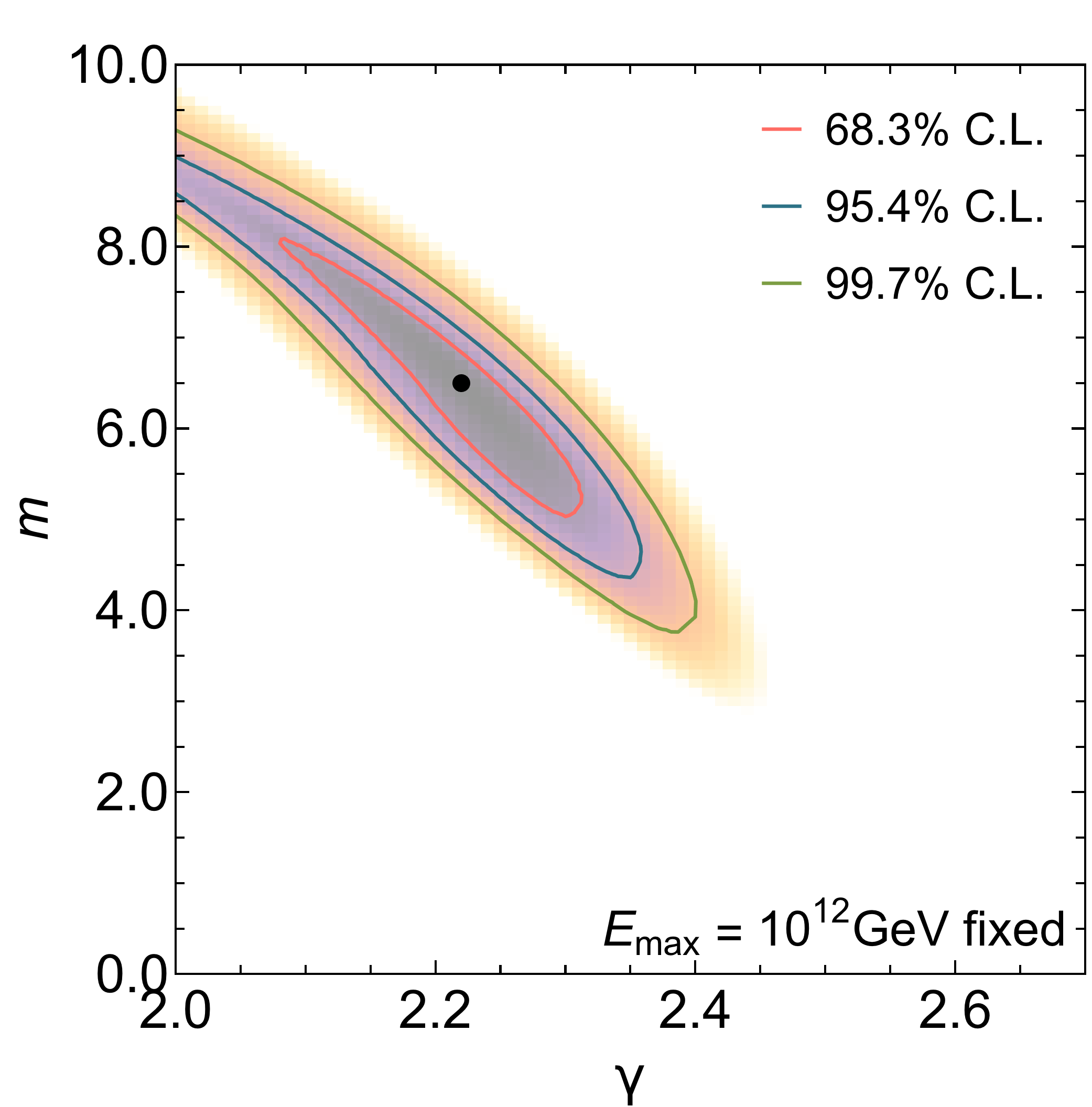} \hspace*{0.3cm} &
\includegraphics[width=0.48\textwidth]{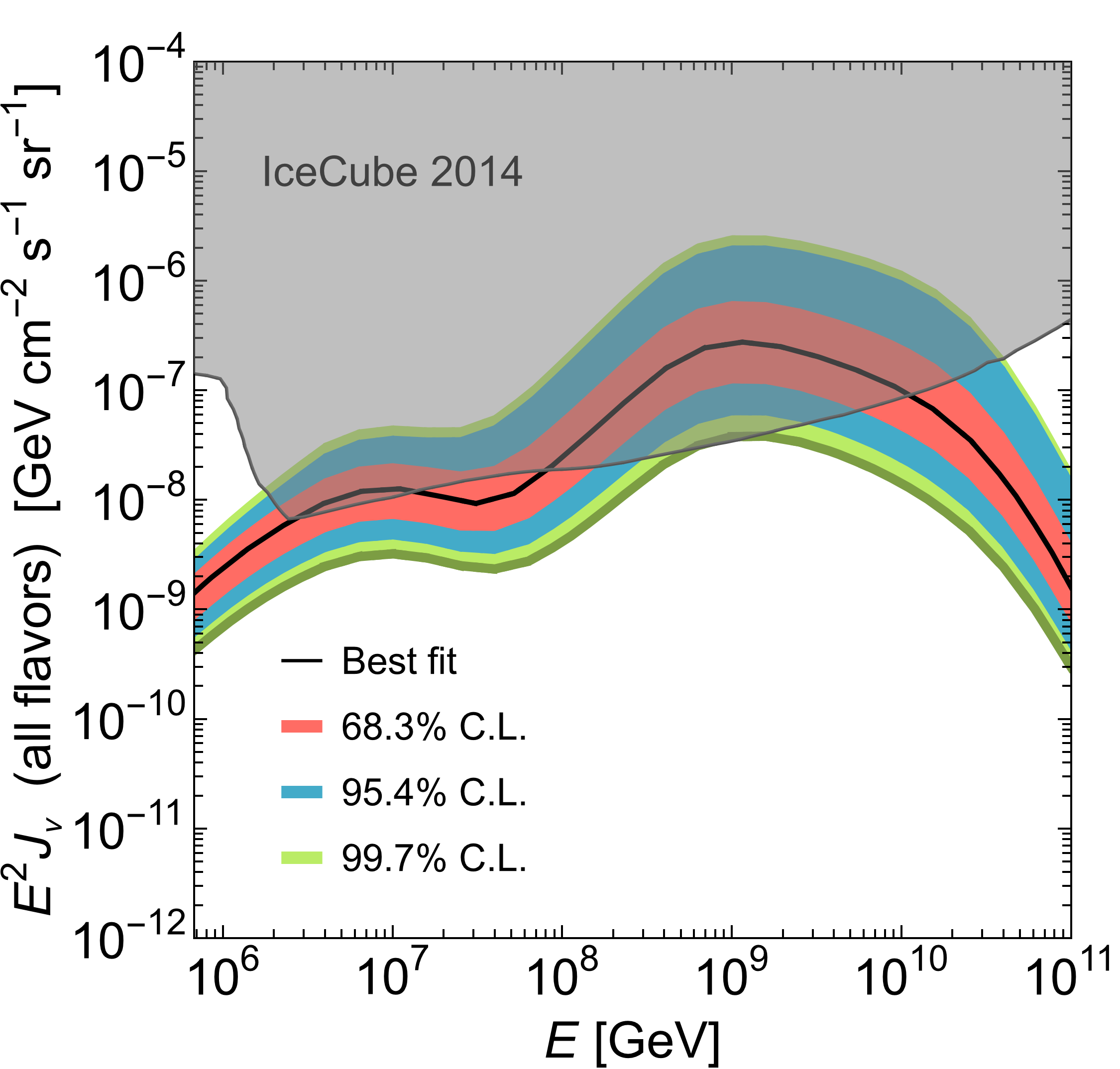}
\end{tabular}
\caption{{\it Left:} Confidence regions of $(\gamma,m)$ found with the TA assumptions of $z_{\mathrm{max}}=2$, $E_{\mathrm{max}} = 10^{12}$ GeV, and the same evolution function used in \citet{Kido:2015}. {\it Right:} Associated neutrino flux. The bands represent the confidence levels as explained in \Sec~\ref{sec.neu}. The IceCube upper limit is from \citet{icecubeupdate}.}
\label{fig.TAresults}
\end{figure*}

Figure~\ref{fig.TAresults}, left panel, shows the allowed region, with the same range for $\gamma$ as in \citet{Kido:2015}. The  $\chi_{\mathrm{min}}^2/{\mathrm{d.o.f.}}$ reported therein is smaller than the one found here, since we do not add the systematic uncertainties related to the event reconstruction  in quadrature to the statistical ones. Moreover, in the present work we use the combined energy spectrum after 7 years of data-taking \citep{Ivanov:2015}, instead of only SD data after 5 years. Nevertheless, we checked that we are able to reproduce the TA results in \citet{Kido:2015} using the 5-year SD data. The remaining differences between the two analyses are related to differences in propagation method and model choices: \citet{Kido:2015} used the transport code \texttt{TransportCR} from \citet{Kalashev:2014xna} and the CIB model by \citet{Kneiske:2002wi}. Regardless, the results of the two analyses are in good agreement.

Figure~\ref{fig.TAresults}, right panel, shows the neutrino flux corresponding to this fit. It has a higher energy cutoff compared to our main result (\Fig~\ref{fig.neulimit}), due to the higher maximal energy assumed. On the other hand, the neutrino flux is higher in our main result because in it injection is continued up to $z=6$.

Table \ref{tab.alt} shows the number of neutrino events at IceCube after 6 years of exposure for different fits. For the 2D fit with TA assumptions (second row), the minimal number of events obtained for a proton flux compatible with TA at 99.7$\%$ C.L. is large enough to reject the proton dip model at 95$\%$ C.L.

%% file: appB.tex
\begin{figure*}[t!]
\centering
\includegraphics[width=0.8\textwidth]{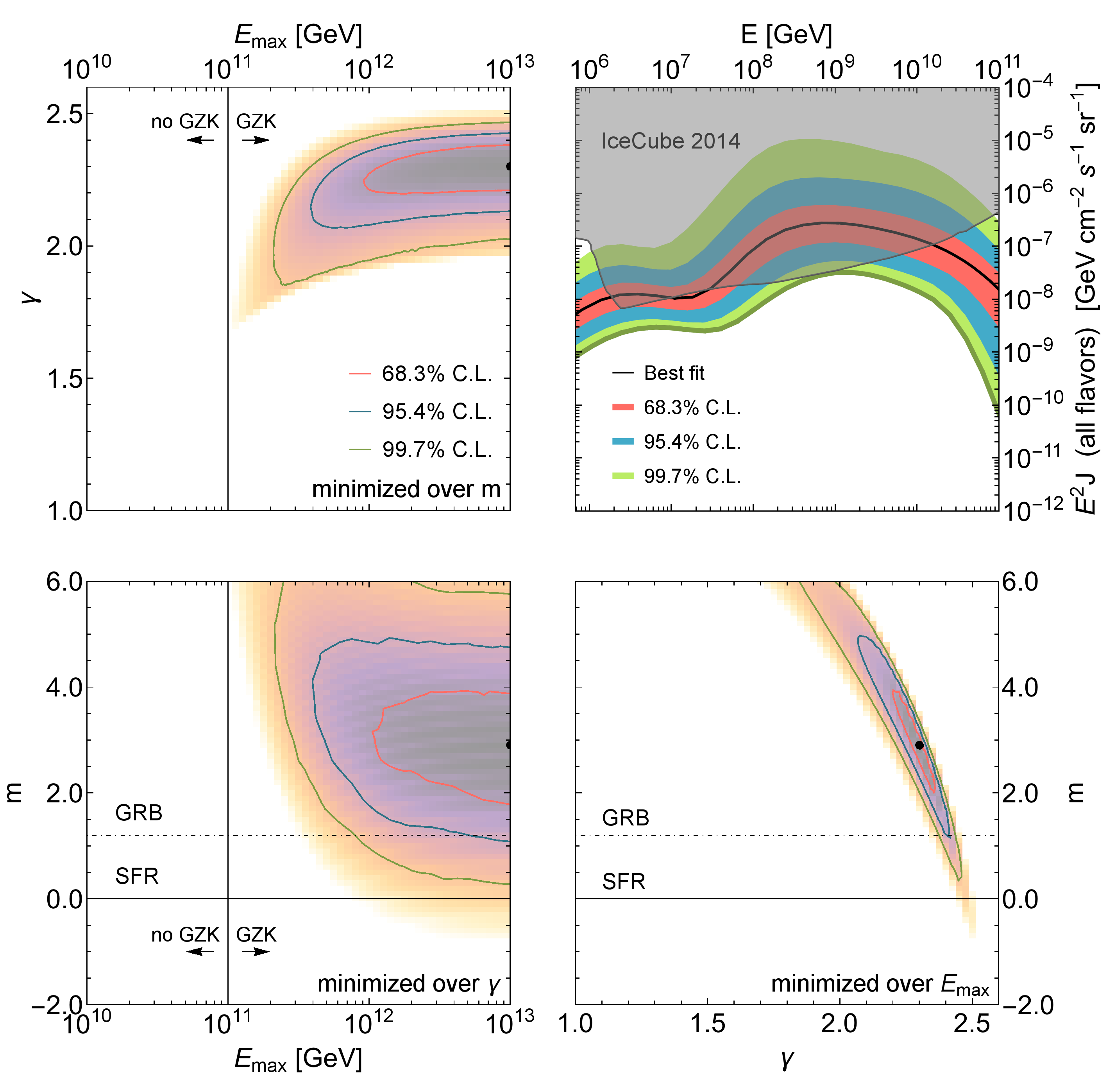}
\caption{Same as Figs.~\ref{fig.3Dscan} and~\ref{fig.neulimit}, but for fixed energy scale ($\delta_E = 0$)
}
\label{fig.3DfixedE}
\end{figure*}

We discuss variations to the model and fitting procedure described in \Sec~\ref{sec.fit} to demonstrate that our conclusions are robust.

\subsection{Effect of the energy scale uncertainty}\label{sec.enescale}
The systematic uncertainty of the energy determination is considered in our procedure as a penalty in the calculation of the $\chi^2$ estimator. Our reference 3D scan presented in \Sec~\ref{sec.fit} features a large pull for the shift $\delta_E$; see Eq.~(\ref{eq.chi}). This is naturally found together with a low value for the maximal energy at the source. A combination of strong source evolution and hard spectrum is then needed to reproduce the low-energy part of the fit region. The corresponding flux of neutrinos exceeds the experimental limit. Here we verify that the minimal neutrino flux is not substantially affected by the extreme value of the shift in the energy scale. This is done by performing a three-dimensional scan over $(\gamma,E_{\mathrm{max}},m)$, fixing $\delta_E = 0$ and minimizing only over the normalization of the flux $f$. The best-fit parameters are reported in Table \ref{tab.alt} (third row).

Figure~\ref{fig.3DfixedE}, left upper panel, shows that low $E_\text{max}$ and low $\gamma$ are not allowed anymore, implying that the region for the allowed source evolution is smaller than in the reference case, and that the GZK cutoff is favored over the source cutoff interpretation. Then, the best-fit neutrino flux will be naturally lower than in the reference case, where the source evolution was found to be considerably higher, and will have a cutoff at higher energies. However, by looking at the number of expected events in Table~\ref{tab.alt} (third row), we can still exclude the proton dip model at the 95\% C.L, confirming that the extreme value of the energy shift found in the reference case does not substantially affect the main conclusion of this work.

\subsection{Effect of an uncorrelated bin-to-bin systematic error}\label{sec.syserror}
The reconstruction of a cosmic ray event introduces a systematic error in the flux, which we have ignored in our main analysis; in \citet{Kido:2015} this contribution, reported as $3\%$, is added in quadrature to the statistical error in each energy bin. Here we test the effect of adding this contribution in our procedure. However, this is a very conservative treatment, as one would expect a strong correlation between bins.

\begin{figure*}[t!]
\centering
\includegraphics[width=0.8\textwidth]{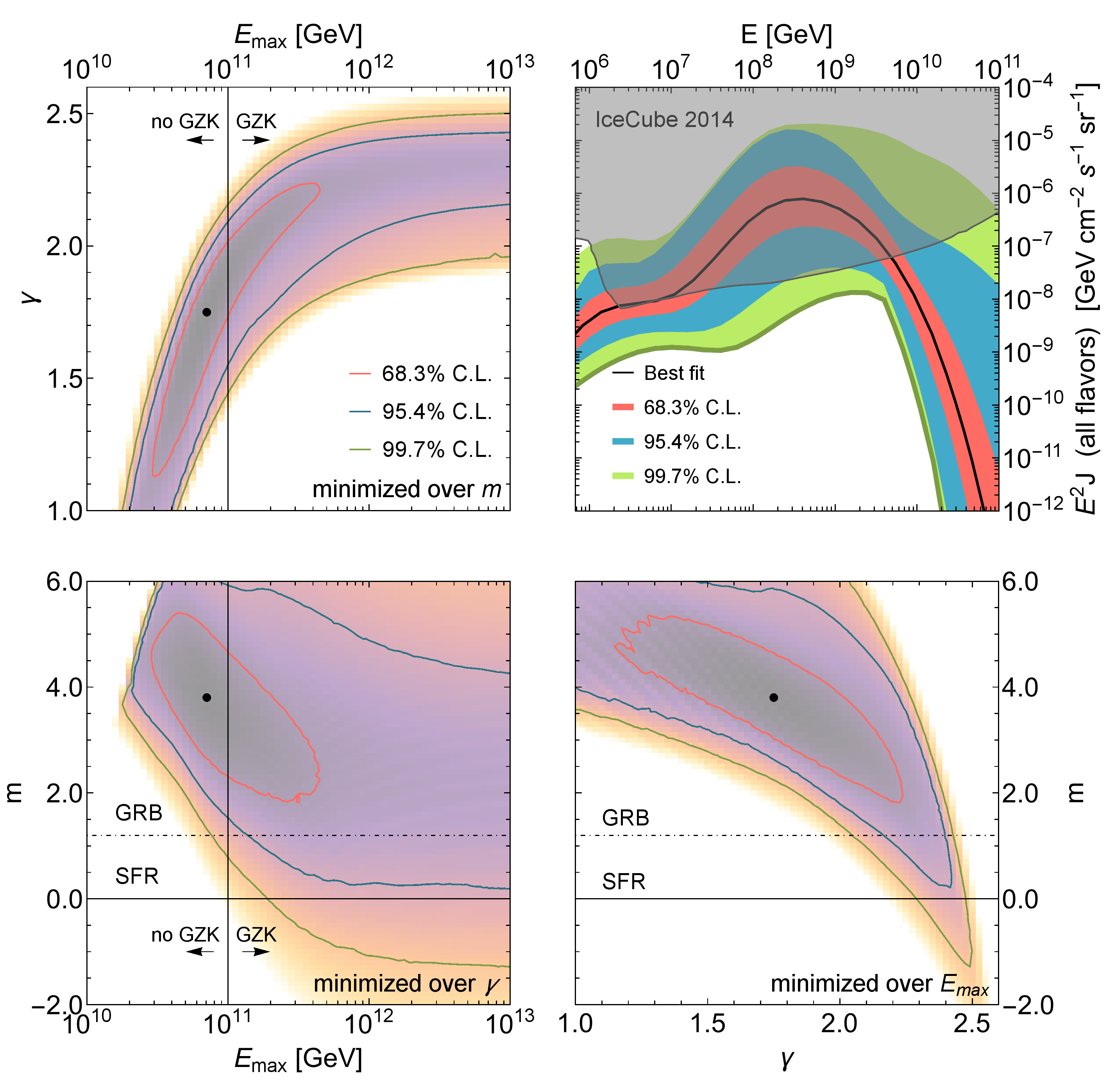}
\caption{Same as Figs.~\ref{fig.3Dscan} and~\ref{fig.neulimit}, but with the 3\% uncorrelated bin-to-bin systematic errors added in quadrature to the statistical errors.
}
\label{fig.3Dsyst}
\end{figure*}

Figure~\ref{fig.3Dsyst} shows the results. The best-fit parameters are reported in Table \ref{tab.alt} (fourth row). The same astrophysical scenario is found as for the reference scan. However, this treatment of the systematics has the consequence of enlarging the experimental error in each energy bin, so that the region of the allowed cosmic ray parameters is larger than in the reference case; the effect is clearly visible mainly in the source evolution. As a consequence, the minimal neutrino flux is now below the IceCube limit and the number of expected events reported in Table~\ref{tab.alt} (fourth row) drops. However, looking at the number of events corresponding to the 95.4\% C.L., there is still tension with IceCube data at more than 95\% C.L. We recall that the 3\% systematic error used here is taken from \citet{Kido:2015}, where the data set used is the SD one, after 5 years of data-taking. Since we use instead the combined spectrum after 7 years, 3\% is probably not properly taking into account the systematic uncertainties. Nevertheless, Fig.~\ref{fig.3Dsyst} shows that this treatment of systematic uncertainties with no bin-to-bin correlations allows a larger degeneracy between source evolution and spectral index, both of which are mainly determined by the lower-energy region of the fit. On the other hand, the cutoff energy $E_\text{max}$ is not affected, since it is mainly determined by the highest-energy region of the fit, where statistical errors exceed the systematic ones. 

\begin{figure*}[t!]
\centering
\includegraphics[width=0.8\textwidth]{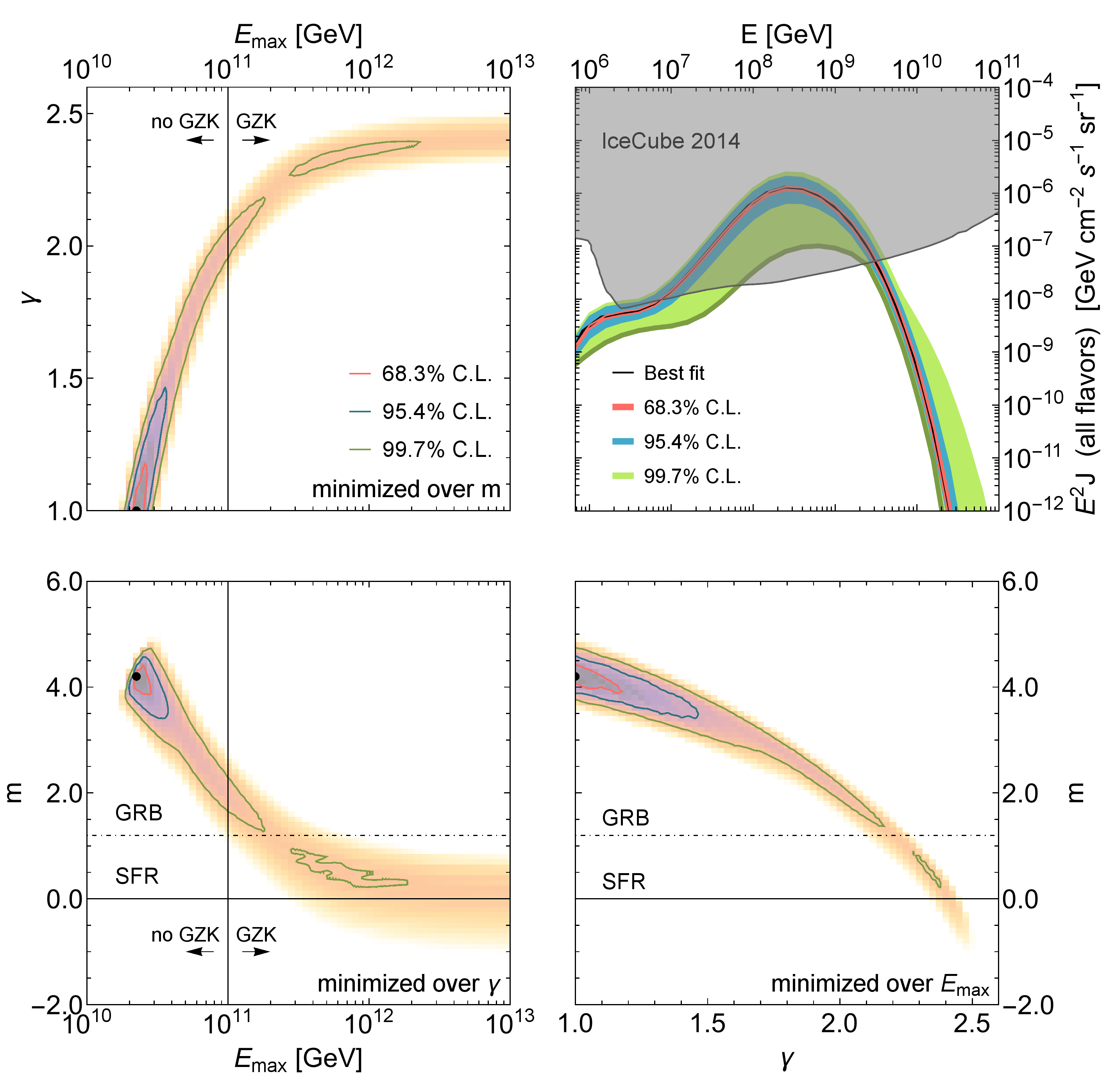}
\caption{Same as Figs.~\ref{fig.3Dscan} and~\ref{fig.neulimit}, but including a penalty for overshooting the flux at low energy.
}
\label{fig.3Dpenalty}
\end{figure*}

\begin{figure*}[t!]
\centering
\includegraphics[width=0.8\textwidth]{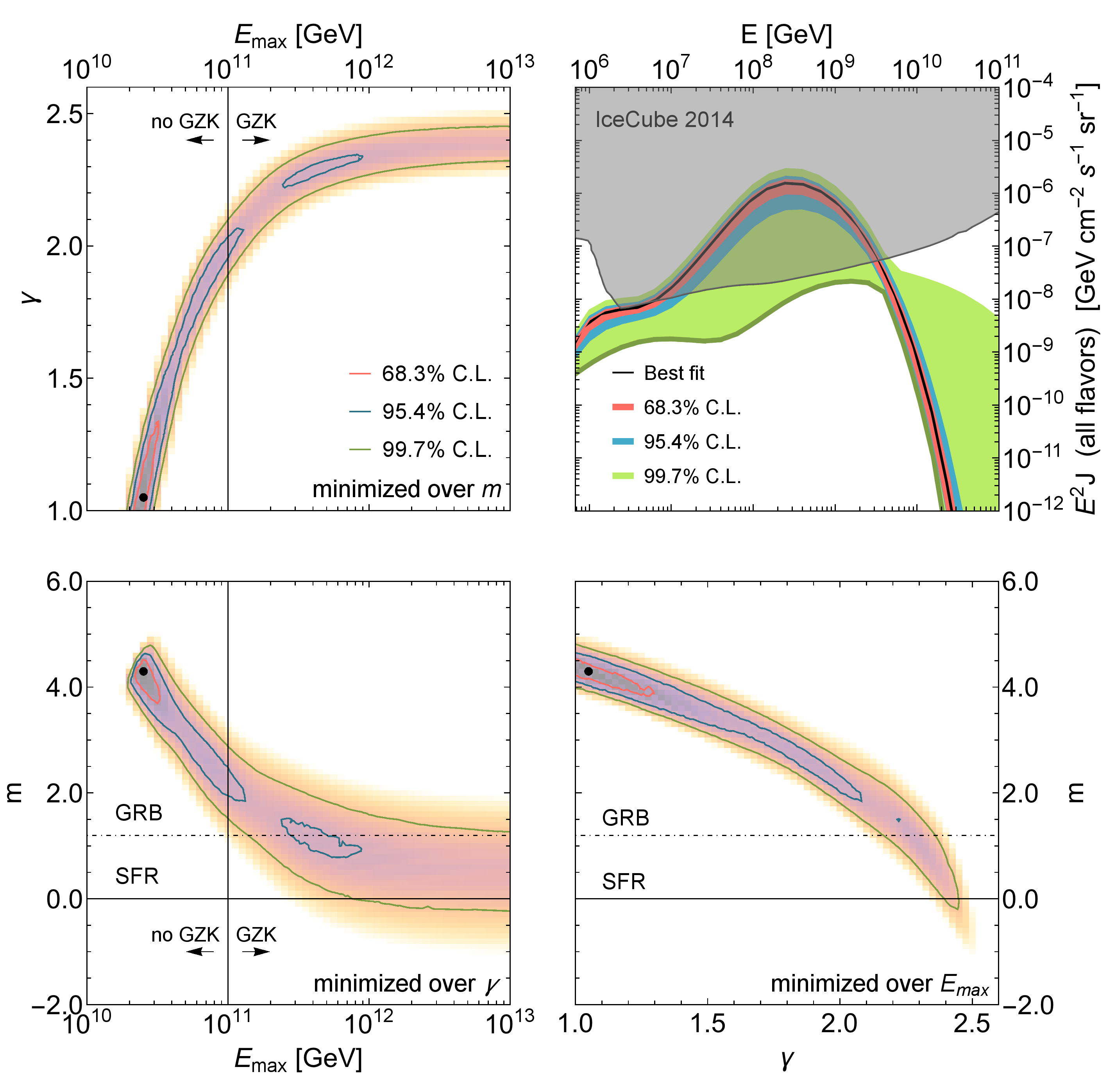}
\caption{Same as Figs.~\ref{fig.3Dscan} and~\ref{fig.neulimit}, but starting the fit from $10^{9}$ GeV.}
\label{fig.Emin90}
\end{figure*}

\begin{figure*}[b!]
\centering
\includegraphics[width=.45\textwidth]{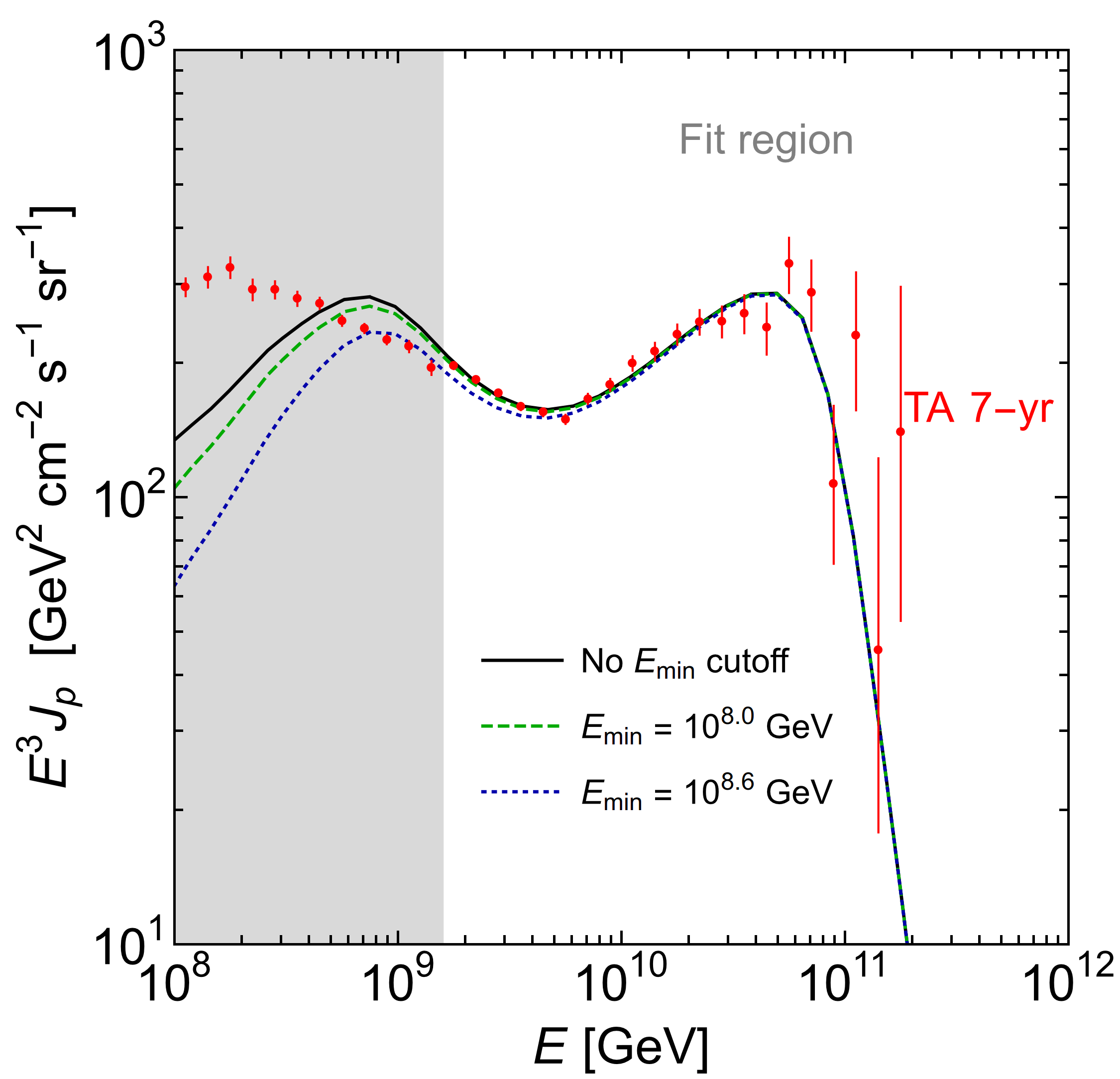}
\caption{Same as \Fig~\ref{fig.protonspectra} for best-fit case of 3D~scan, and several assumptions for a minimal injection energy cutoff.}
\label{fig.protonspectra_Emin}
\end{figure*}

\subsection{Overshoot penalty at low energies}\label{sec.oversh}
The strong source evolution found in our analysis (and also in the TA analysis) has the consequence of overshooting the flux at the lowest energies. In order to avoid that, a penalty can be included in the fit: the $\chi^2$ corresponding to the data points in the gray-shaded region in \fig~\ref{fig.protonspectra} is added to the total $\chi^2$ if the flux overshoots the measurement. The best-fit parameters are reported in Table \ref{tab.alt} (fifth row). Figure~\ref{fig.3Dpenalty} shows the results.
The allowed regions are narrower compared to the 3D scan without the penalty. The best-fit value of $\delta_E$ is even more extreme and allows a small maximal energy. The source evolution is similar, but it is now associated to harder spectra. The best-fit neutrino flux is similar, but the lowest flux is now well above the IceCube limit in the range $\sim 7\times 10^{7} - 2\times 10^{9}$ GeV. This is reflected in higher event numbers in Table~\ref{tab.alt} (fifth row). In this scenario, the proton dip model can be excluded at $>99\%$ C.L.

\subsection{Effect of changing the starting energy of the fit}\label{sec.startingene}
We performed our reference fit above $10^{9.2}$ GeV in order to directly compare it to the fit performed by \citet{Kido:2015}, who chose the same starting energy, while using the SD-only energy spectrum. Though we used instead the combined spectrum, this is dominated by SD data, which further justifies our choice of starting energy. We have explored the effect of using a lower starting energy in the fit. Since the proton spectrum is affected by magnetic fields in the range $10^8-10^9$ GeV \citep{Globus:2007bi,Kotera:2007ca}, we tested $10^9$ GeV as the starting energy. Doing so, we added two more TA data points to the fit. The best-fit parameters are reported in Table \ref{tab.alt} (sixth row). Figure~\ref{fig.Emin90} shows the results. Previsibly, the best-fit parameter values are very similar to the case where a penalty for overshooting at the lowest energies was included. The 95.4$\%$ C.L.~region in this case is very similar to the 99.7$\%$ C.L.~of that case, while, within the 99.7$\%$ C.L., low values for the source evolution are allowed. As a consequence, the lowest neutrino flux is below the IceCube limit and the corresponding number of expected neutrino events drops, as can be seen in Table~\ref{tab.alt} (sixth row). We also notice that by including these two data points the fit gets worse. The reason may be that the SD data set dominates in the energy range above $10^{9.2}$ GeV, while other data sets significantly contribute at lower energies -- which may lead to some systematics not taken into account by our procedure.

\subsection{Effect of minimal injection energy}\label{sec.mininj}
The overshoot at low energies may, however, be physical. Several mechanisms have been considered to reduce the flux at low energies, as already reported in Sec.~\ref{sec.fit}. A detailed discussion of the flux at the lowest energies would go beyond the scope of this paper. However, we verify here that the overshooting of the flux at low energy can be avoided by multiplying \eq~(\ref{equ:injflux}) by the factor $\exp(-E_{\mathrm{min}}/E)$, where we use $E_{\mathrm{min}}=10^8,10^{8.6}$ GeV (in the co-moving frame). Figure~\ref{fig.protonspectra_Emin} shows the results.

\begin{figure*}[t!]
\centering
\includegraphics[width=0.8\textwidth]{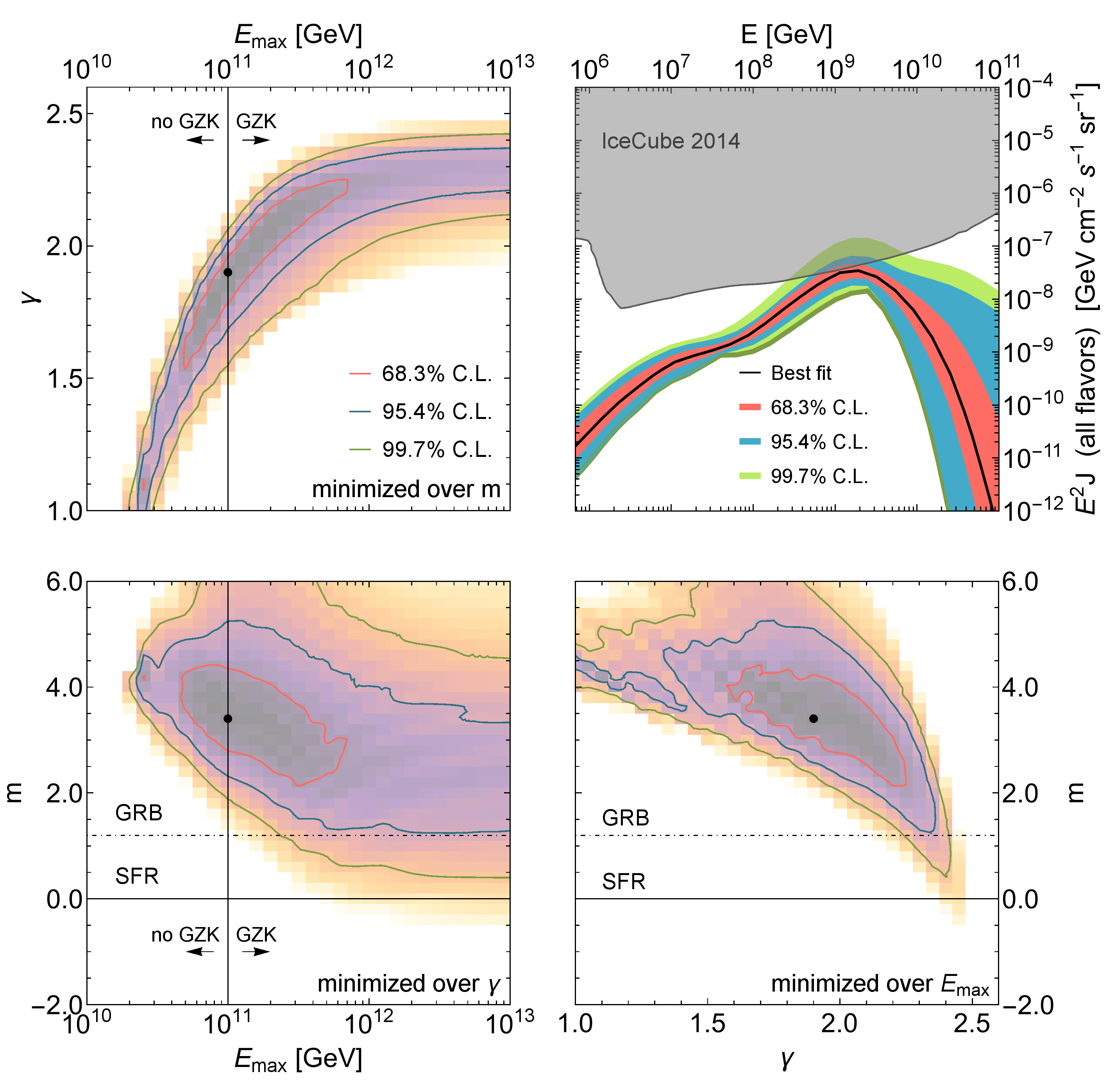}
\caption{Same as Figs.~\ref{fig.3Dscan} and~\ref{fig.neulimit}, but with no proton injection for $z>1$.
}
\label{fig.3Dz1}
\end{figure*}

\subsection{Maximal redshift injection cutoff}\label{sec.maxredshift}
We performed our 3D fit in \Sec~\ref{sec.fit} above $10^{9.2}$ GeV, following \citet{Kido:2015}. In this range, the contribution to the cosmic ray flux is dominated by sources with $z<1$, corresponding to the maximal interaction length at this minimal energy. We repeat here the fit procedure considering a distribution of sources up to $z=1$, instead of $z = 6$. The best-fit parameters are reported in Table \ref{tab.alt} (seventh row). 
The overshooting of the flux at low energies in this case is avoided by construction (as the low energies are dominated by higher redshifts), and the natural consequence is to make the allowed regions shift towards larger values of the spectral index and maximal energy. Figure~\ref{fig.3Dz1} shows that the allowed parameter regions in this fit are not too different from those obtained in \Sec~\ref{sec.fit}.  However, the neutrino results are strikingly different. The flux corresponding to the best fit is now at the level of the IceCube limit around $10^9$ GeV. Table~\ref{tab.alt} (seventh row) shows that the corresponding number of events drops appreciably. This means that the UHECR spectrum does not have enough power to constrain the parameter space at large redshifts. On the other hand, the neutrino flux is strongly affected by large redshifts. Thus, while the UHECR spectrum alone could lead to degeneracy -- especially in the determination of the source evolution -- the comparison between the predicted cosmogenic neutrino flux and its experimental limits can probe source evolution well beyond the local universe.
 
We note that we observe a similar effect if the fit energy range starting at $10^9$ GeV is combined with the uncorrelated bin-to-bin systematic errors. In that case,  a larger degeneracy in the parameter space with respect to the reference case is observed, and negative source evolution (corresponding to a reduction of the injection) is approached at the 99.7\% C.L. Consequently, the associated number of neutrino events drops.